# Anisotropic Elastic Properties of Flexible Metal–Organic Frameworks: How Soft are Soft Porous Crystals?

Aurélie U. Ortiz[1], Anne Boutin[2], Alain H. Fuchs[1], and François-Xavier Coudert[1]*

[1] *CNRS and Chimie ParisTech, 11 rue Pierre et Marie Curie, 75005 Paris, France*

[2] *Département de Chimie, École Normale Supérieure, CNRS-ENS-UPMC, 24 rue Lhomond, 75005 Paris, France*

(Dated: December 11, 2014)

We performed ab initio calculations of the elastic constants of five flexible metal–organic frameworks: MIL-53(Al), MIL-53(Ga), MIL-47 and the square and lozenge structures of DMOF-1. Tensorial analysis of the elastic constants reveal a highly anisotropic elastic behavior, some deformation directions exhibiting very low Young's modulus and shear modulus. This anisotropy can reach a 400:1 ratio between the most rigid and weakest directions, in stark contrast with the case of non-flexible MOFs such as MOF-5 and ZIF-8. In addition, we show that flexible MOFs can display extremely large negative linear compressibility (NLC). These results uncover the microscopic roots of stimuli-induced structural transitions in flexible MOFs, by linking the local elastic behavior of the material and its multistability.

Much attention has recently been focused on a fascinating subclass of metal–organic frameworks (MOFs) that behave in a remarkable stimuli-responsive fashion.[1] The number of reported syntheses of such flexible MOFs, also called Soft Porous Crystals (SPCs),[2] is rapidly growing and they are promising for practical applications, such as gas capture, purification and fluid separation. These materials feature dynamic crystalline frameworks displaying reversible structural deformations of large amplitude under a number of external physical constraints such as guest adsorption, temperature or mechanical pressure. The latter was the most recently demonstrated of the possible stimuli of SPCs, with a clear observation of stress-induced reversible crystal-to-crystal structural transitions in MIL-53(Cr).[3] Moreover, the stress induced on the host framework by guest adsorption also plays a big role in the structural transitions observed upon fluid adsorption.[4] It is thus a key quantity in the description of the flexibility of MOFs.

In spite of this, there is a major lack of data about the fundamental mechanical characteristics of SPCs that link stress $\sigma$ and framework deformation $\epsilon$ through Hooke's law:

$$\sigma_{ij} = C_{ijkl}\,\epsilon_{kl} \tag{1}$$

with C the elasticity tensor (or stiffness tensor) of the material. The only mechanical property of SPCs that have been reported so far is their bulk modulus. It was estimated for MIL-53(Al) from volumetric data on compression with mercury[5], and confirmed by means of molecular simulation on the similar MIL-53(Cr) material.[6] Moreover, it was measured for $NH_2$-MIL-53(In) by powder X-ray diffraction upon compression.[7] However, these measurement of the scalar bulk modulus fail to account for the tensorial nature of the generalized Hooke's law (Eq. 1), and other crucial elastic properties of SPCs, like their Young's modulus, shear modulus or Poisson's ratio, have not yet been investigated. In contrast, the full elastic constants tensors of a few non-compliant metal–organic frameworks, like MOF-5[8] and ZIF-8,[9] have been reported[10, 11] and were shown to yield much information about the elastic behavior and structural stability of MOFs, a critical factor for any practical application. In this letter, we elucidate the anisotropic elastic properties of Soft Porous Crystals and shed light onto the microscopic manifestations at the origin of flexibility in these materials, contrasting them with the properties previously observed for non-compliant frameworks of a similar chemical nature.[10, 11]

In order to characterize the elastic properties of SPCs, we have calculated the full elastic constants tensor of 5 different flexible MOF structures. Two of the materials chosen were from the MIL-53 family of materials (Fig. 1): MIL-53(Al) and MIL-53(Ga)-lp. Their frameworks are made of parallel one-dimensional M(OH) chains (M = $Al^{3+}$, $Ga^{3+}$) linked together by 1,4-benzenedicarboxylate linkers to form linear diamond-shaped channels that are wide enough to accommodate small guest molecules. Upon guest adsorption[1, 12] or mechanical pressure,[5] this structure may oscillate (or "breathe") between two distinct conformations called the large-pore phase ("lp", which we studied) and the narrow-pore phase ("np"), which have a remarkable difference in cell volume of up to 40%. We also studied MIL-47, a similar vanadium-based material built from VO chains linked by the same organic linker, which has never been observed to breathe under adsorption but was recently shown to contract to a narrow pore phase under pressure.[13] Finally, we also included two known structures of zinc-based flexible MOF, the DMOF-1: a full expanded structure with square one-dimensional channels built on DABCO (1,4-diazabicyclo[2.2.2]octane) pillars, and its contracted version with lozenge-shaped channels. We also compared the behavior of these flexible frameworks with two other MOF structures whose elastic constants have been published: MOF-5 and ZIF-8. The structures of these materials are depicted in Fig. 1.

The single crystal elastic constants of these five structures were calculated using *ab initio* quantum mechanical calculations[14] in the density functional theory approach with localized basis sets (CRYSTAL09 code[15]). We used the B3LYP hybrid exchange–correlation functional,[16] whose accuracy for the calculations of MOF structures,[17] energies [18] and elastic constants[11, 19] is now well established. The elastic constants (elements of the stiffness tensor) thus obtained are reported in Table I; because all the materials chosen have an orthorhombic unit cell, each has 9 independant stiffness constants. In each case, they verify the Born stability condition, namely that the symmetric stiffness tensor is positive-definite. From the stiffness constants, a full tensorial analysis was performed and key quantities were derived that characterize the mechanical behavior of the crystal in the elastic regime (schematized on Fig. 2):

- Young's modulus, $E(\mathbf{u})$, characterizes the uniaxial stiffness of the material in the direction of unit vector $\mathbf{u}$.

- The linear compressibility, $\beta(\mathbf{u})$, quantifies the deformation in direction $\mathbf{a}$ as a response to isostatic compression.

- The shear modulus, $G(\mathbf{u}, \mathbf{n})$, characterizes the resistance to shearing of the plane normal to $\mathbf{n}$ the $\mathbf{u}$ direction.

- Poisson's ratio, $v(\mathbf{u}, \mathbf{v})$, is the ratio of transverse strain in direction $\mathbf{v}$ to axial strain in direction $\mathbf{u}$, when uniaxial stress is applied.

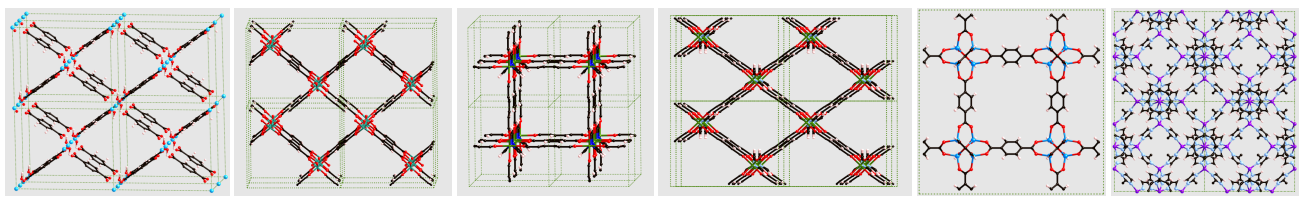

Figure 1. From left to right: MIL-53(Al) lp structure, MIL-47, DMOF1-sq, DMOF1-loz, MOF-5, and ZIF-8.

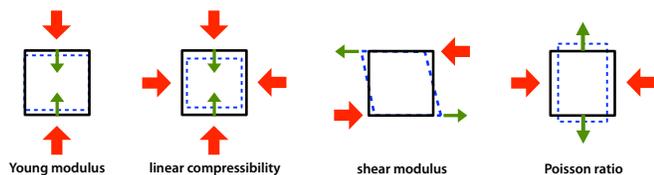

Figure 2. Scheme of the directional elastic properties calculated in this work. For each, large red arrows represent the direction of applied stress and smaller green arrows the direction along which the resulting strain is measured.

| $C_{ij}$ | MIL-53(Al) | MIL-53(Ga) | MIL-47 | DMOF-1 | DMOF-1 |
|---|---|---|---|---|---|
| (in GPa) | lp | lp | | loz | sq |
| $C_{11}$ | 90.85 | 112.32 | 40.69 | 57.15 | 35.33 |
| $C_{22}$ | 65.56 | 56.66 | 62.60 | 35.59 | 58.20 |
| $C_{33}$ | 33.33 | 18.52 | 36.15 | 17.68 | 58.45 |
| $C_{44}$ | 7.24 | 5.48 | 50.83 | 0.62 | 0.11 |
| $C_{55}$ | 39.52 | 21.71 | 7.76 | 16.39 | 0.44 |
| $C_{66}$ | 8.27 | 6.64 | 9.30 | 0.69 | 0.28 |
| $C_{12}$ | 20.41 | 22.87 | 12.58 | 9.85 | 7.32 |
| $C_{13}$ | 54.28 | 45.35 | 9.28 | 31.43 | 7.55 |
| $C_{23}$ | 12.36 | 10.86 | 46.98 | 5.47 | 11.68 |

Table I. Stiffness constants $C_{ij}$ in Voigt notation for the five MOFs studied.

The directional dependence of the above-listed properties can be calculated from the 4th order compliance tensor $\mathsf{S}$, which is the inverse of the stiffness tensor $\mathsf{C}$, by applying to it a rotation mapping the $x$ and $y$ axes onto the directions of $\mathbf{u}$ and $\mathbf{v}$ to obtain a rotated tensor $\mathsf{S}'$. Young's modulus, being the axial response to a purely axial stress, is calculated as

$$E(\mathbf{u}) = \frac{1}{S'_{1111}(\mathbf{u})} = \frac{1}{u_i u_j u_k u_l S_{ijkl}} \quad (2)$$

Other properties can be similarly expressed as functions involving the components of tensor $\mathsf{S}$ and unit vectors $\mathbf{u}$ and $\mathbf{v}$:[20]

$$\beta(\mathbf{u}) = u_i u_j S_{ijkk} \quad (3)$$

$$G(\mathbf{u}, \mathbf{v}) = \left( u_i v_j u_k v_l S_{ijkl} \right)^{-1} \quad (4)$$

$$v(\mathbf{u}, \mathbf{v}) = -\frac{u_i u_j v_k v_l S_{ijkl}}{u_i u_j u_k u_l S_{ijkl}} \quad (5)$$

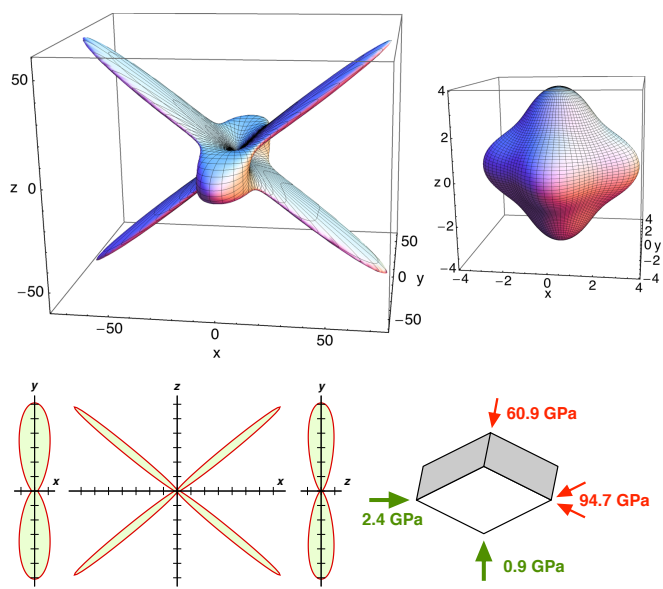

Figure 3. Directional Young modulus for MIL-53(Al)-lp and ZIF-8 represented as 3D surfaces, with axes tick labels in GPa (top left and right, respectively); projection in the $xy$, $xz$ and $zy$ planes (bottom left; one tick is 10 GPa). A scheme showing the stiffest and weakest directions of the lozenge-shaped pore is presented at the bottom right.

We first focus on Young's modulus, which is plotted for MIL-53(Al)-lp on Fig. 3 for the (001), (010) and (100) planes. A 3D surface representation is also provided, which corresponds to a spherical plot of $E(\mathbf{u})$ depending on the direction of unit vec-tor **u**. It is clearly visible that Young's modulus for this material is very anisotropic, with high-value lobes in the $y$ direction (60.9 GPa) as well as along two directions in the $xz$ plane (94.7 GPa). The first one corresponds to the axis of the channel, and thus to compression of the inorganic Al(OH) chain, which is expected to be quite resistant to compression. The second one, which makes an angle of ±38° with the $x$ axis, corresponds to compression along the organic linkers, which also explains the stiffness. In other directions, such as the $x$ and $z$ crystallographic axes, it is very low (2.4 GPa and 0.9 GPa respectively). These two directions, which are the principal axes of the lozenge-shaped channel, correspond to the "breathing" mode of deformation for the solid: while pressing on the $x$ axis, the lozenge can deform by elongating along the $z$ axis with the length of all linkers staying constant. This anisotropy, characterized as the ratio $A_E = E_{max}/E_{min}$ of the Young's modulus in the stiffest direction to the minimal Young's modulus, has an extremely high value of 105.

| property | $E_{min}$ (GPa) | $E_{max}$ (GPa) | $A_E$ | $G_{min}$ (GPa) | $G_{max}$ (GPa) | $A_G$ | $\beta_x$ (TPa$^{-1}$) | $\beta_y$ (TPa$^{-1}$) | $\beta_z$ (TPa$^{-1}$) | $\nu_{min}$ | $\nu_{max}$ |
|---|---|---|---|---|---|---|---|---|---|---|---|
| MIL-53(Al) lp | 0.90 | 94.7 | 105 | 0.35 | 39.5 | 112 | −257 | 11 | 445 | −2.4 | 1.9 |
| MIL-53(Ga) lp | 0.16 | 69.7 | 444 | 0.08 | 21.7 | 270 | −1441 | −98 | 3640 | −6.2 | 2.9 |
| MIL-47 | 0.9 | 96.6 | 108 | 0.29 | 50.8 | 175 | 22 | −201 | 283 | −1.5 | 2.2 |
| DMOF-1 loz | 0.39 | 46.3 | 119 | 0.16 | 16.4 | 102 | −623 | 23 | 1158 | −0.4 | 3.2 |
| DMOF-1 sq | 0.45 | 55.0 | 123 | 0.11 | 18.4 | 165 | 23 | 12 | 12 | 0.00 | 1.0 |
| MOF-5 (ref. 10) | 9.5 | 19.7 | 2.1 | 3.4 | 7.5 | 2.2 | 20 | 20 | 20 | 0.03 | 0.67 |
| ZIF-8 (ref. 11) | 2.7 | 3.9 | 1.4 | 0.94 | 1.4 | 1.4 | 36 | 36 | 36 | 0.33 | 0.57 |

Table II. Minimal and maximal values as well as anisotropy of Young's modulus, shear modulus, linear compressibility and Poisson's ratio for the MOFs studied. Anisotropy of $X$ is denoted by $A_X = X_{max}/X_{min}$.

Table II reports the values of maximal and minimal Young's modulus calculated for all five flexible MOFs studied, as well as for MOF-5 and ZIF-8. The strong anisotropy observed on MIL-53(Al)-lp is clearly observed for all flexible materials. It is further confirmed by the full surface representations (see Supplemental Information); in particular, in all cases the stiff directions correspond either to inorganic chains or organic linkers, while the softer directions correspond to the "breathing" deformation modes. This is in sharp contrast with behaviour of Young's modulus for MOF-5 and ZIF-8 whose asymmetry is two orders of magnitude lower. It can also be seen that the values of Young's modulus for flexible MOFs in their direction of lowest rigidity, all below 1 GPa, are much lower than that of other MOFs and zeolites, which are typically in the range of 1 to 10 GPa. Thus, we conclude that the presence of directions of very low Young's modulus, and thus the high anisotropy of Young's modulus, are revealing signatures of the structural flexibility of the soft porous crystals. This is similar to the way low-frequency vibration modes of a molecular structure are indicators of its conformational flexibility.

We also analyzed the shear modulus $G$ of all structures studied. The 3D representation of shear modulus is harder than for Young's modulus, since $G(\mathbf{u}, \mathbf{n})$ is a function of two unit vec-tors rather than one. We thus characterized the shear modulus by its maximal and minimal values as a function of direction $(\theta, \phi)$ with respect to the third parameter $\chi$. Without going into details, we see again from Table II and Fig. 4 that the shear modulus anisotropy is much higher for the soft porous crystals than for the other MOFs. Moreover, the shear stresses corresponding to the weakest modulus are located in the cross-section of the channels (lozenge or square), and along the organic linkers. Like in the case of Young's modulus, they thus correspond to the breathing deformation and are a clear indicator of structural flexibility. In particular, the directions along which shearing is most easy in the MIL-53(Al)-lp framework are the same as those of the layer-by-layer shearing mechanism as independently predicted from elastic compatibility equations[21] and observed in molecular simulations.[22] This justifies the approximation made in our earlier work on a simple model describing the structural transition at the level of the crystal, assuming that the main deformation mode of MIL-53 keeps the linker length constant but shears the pore channel.[21]

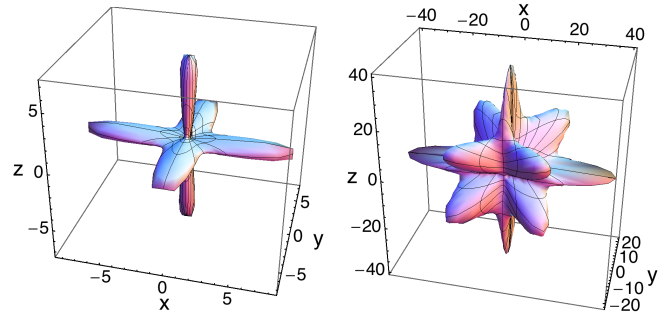

Figure 4. 3D representations of the shear modulus for MIL-53(Al)-lp. Left: minimal shear modulus $G_{min}(\theta, \phi) = \min_\chi G(\theta, \phi, \chi)$; right: maximal shear modulus $G_{max}(\theta, \phi) = \max_\chi G(\theta, \phi, \chi)$. Axes tick labels are in GPa.

We now turn to the linear compressibility $\beta$ of these frameworks. The 3D surface representation of $\beta$ is presented in Fig. 5 for MIL-53(Al)-lp, DMOF1-loz and DMOF1-sq. Again, MIL-53(Al)-lp exhibits a large anisotropy in its linear compressibil-ity, with a positive lobe along the $z$ axis, a negative lobe along the $x$ axis and a much smaller positive value along the $y$ axis (which is the channel axis). Indeed, as isostatic pressure is applied to the MIL-53 framework, the overall effect is a contraction of the material, but it needs not be isotropic. First, the inorganic chain in the $y$ axis being quite rigid, it deforms very little, with $\beta_y = 11$ TPa$^{-1}$, which is a typical order of magnitude for inorganic solids. Secondly, the overall narrowing of the pore implies a contraction in the $z$ direction, and hence a positive linear compressibility (445 TPa$^{-1}$), but a simultaneous expansion of the $x$ axis to keep the lozenge side length constant. This results in a negative linear compressibility of $\beta_x = -257$ TPa$^{-1}$. The same effect is observed on all the flexible MOF structures, with the exception of the square DMOF-1 framework. There, the symmetry of the framework in the plane perpendicular to the channels means that purely isostatic compression cannot trigger the breathing of the framework, thus yielding a positive linear compressibility in all directions, with a slight asymmetry between the channel axis ($\beta_y = 27$ TPa$^{-1}$) and its perpendicular plane ($\beta_x = \beta_z = 11$ TPa$^{-1}$). MOF-5 and ZIF-8 belong to the cubic crystal system, and thus display fully isotropic linear compressibility, $\beta(\mathbf{u}) = 1/(C_{11} + 2C_{12})$. As a conclusion, we find that negative linear compressibility, while it is a telltale sign of structural flexibility, may not be observed for all structures of soft porous crystals because of symmetry. However, when it is present, the extent of the negative linear compressibility is one order of magnitude higher than what may be observed in inorganic crystals, where the current "record holder" is $Ag_3[Co(CN)_6]$, with a value of $-75$ TPa$^{-1}$.[23, 24] In fact, the soft porous crystals with their Erector-like compliant frameworks, are archetypal examples of frameworks showing extreme linear compressibility, including both positive and negative lobes.

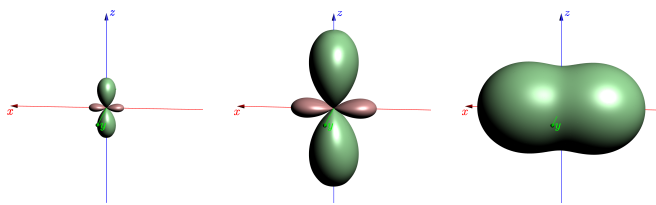

Figure 5. 3D surface representation of the Linear Compressibility of MIL-53(Al)-lp, DMOF1-loz and DMOF1-sq. Positive LC is indicated as green, negative LC in red. The leftmost two are at the same scale (axis length is 1500 TPa$^{-1}$) while the last one is enlarged 50 times for clarity (axis length is 30 TPa$^{-1}$).

Lastly, we look at the bulk modulus of the Soft Porous Crystals. While for crystals in the cubic system there is a direct and unique way to calculate the scalar bulk modulus from the elastic constants, in the three different methods have been proposed for its calculation: the Voigt averaging assumes a uniform strain, the Reuss averaging assumes uniform stress, and the Hill scheme is the geometric average of the previous two. Because the SPCs are highly anisotropic, all three schemes give very different values that fall in the range of 1 to 20 GPa for the solids presented here, indicating that the so-called Soft Porous Crystals may not be so soft overall, at least when it comes to their average elastic properties. In particular, the bulk modulus of MIL-53(Cr)-lp was estimated from mercury compression experiments at around 2 GPa,[5] falling in this range of values.[25]

It is worth noting that, while the tensorial analysis of the elastic tensors of SPCs can reveal key characteristics of their mechanical behavior, its validity is limited to the region of elastic behavior around the relaxed structure. All deformations performed in the calculations herein reported, which correspond to strains of up to ±0.01, fall in this elastic region. However, the full extent of the elastic region is yet to be characterized, as does the behavior of the material for deformations outside the elastic domain. The latter may play an important role in the stimuli-induced structural transitions of SPCs, due to the complexity of their framework and the existence of very soft deformation modes. Work is under way to address these issues by calculating higher-order elastic constants and energy profiles for larger deformations.

In summary, we predicted the mechanical properties of 5 Soft Porous Crystals using quantum mechanical calculations and showed that the framework flexibility and existence of structural transition are clearly visible in their local elastic properties. In particular, the existence of deformation modes of very low rigidity is clearly seen from both their Young's modulus and shear modulus, and contrast sharply with other nonflexible MOFs. We also showed that many of these structures present anomalous elastic behaviour, with both negative Poisson's ration and extremely high negative linear compressibility. Thus, while the elastic behaviour of soft porous crystals is very complex, a full tensorial analysis reveals the key mechanical features of their flexibility and opens up new opportunities for better understanding and tuning their mechanical properties.

The authors thank Anthony K. Cheetham, Arnaud Marmier and Alexander V. Neimark for discussions and insightful comments, and Frédéric Labat for his help with CRYSTAL09 calculations. Funding from the Agence Nationale de la Recherche under project "SOFT-CRYSTAB" (ANR-2010-BLAN-0822) is acknowledged.

---

* fx.coudert@chimie-paristech.fr